\input article.sty\relax
\@addtoreset{equation}{section}
\newcounter{subeqn}

\def\zeqn{\setcounter{subeqn}{0}}
\documentstyle[dina4,12pt]{myart}
\newcounter{subeq}

\newcommand{\subeqno}{\stepcounter{subeq} \addtocounter{equation}{-1}}

\newcommand{\subeqres}{\setcounter{subeq}{0}}
\newcommand{\eqnoinc}{\addtocounter{equation}{1}}
\newcommand{\be}{\begin{equation}}
\newcommand{\ee}{\end{equation}}
\newcommand{\bea}{\begin{eqnarray}}
\newcommand{\eea}{\end{eqnarray}}
\newcommand{\beano}{\begin{eqnarray*}}
\newcommand{\eeano}{\end{eqnarray*}}
\newcommand{\bfr}{\begin{flushright}}
\newcommand{\efr}{\end{flushright}}

\newcommand{\erfc}{{\rm erfc}}
\begin{document}
\title{Microscopic analysis of
 two-body correlations in light nuclei\footnote{supported
 by DFG, BMFT and GSI Darmstadt}}
\author{A. Peter, W. Cassing, J. M. H\"auser and A. Pfitzner$^1$ \\
Institut f\"ur Theoretische Physik, Universit\"at Gie\ss en \\
35392 Gie\ss en, Germany \\
$^1$ Forschungszentrum Rossendorf, 01314 Dresden, Germany }
\maketitle
\begin{abstract}
Within a nonperturbative dynamical two-body approach
- based on coupled equations of motion for the
one-body density matrix and the two-body correlation function  -
we study the distribution of occupation numbers in a correlated system
close to the groundstate, the relaxation of single-particle excitations
and the damping of collective modes.  For this purpose the nonlinear
equations of motion are solved numerically within a finite oscillator basis
for the first time adopting short-range repulsive and long-range
attractive two-body forces. We find in all cases that
the formation of long- and short-range correlations and their mixing is
related to the long- and short-range part of the nucleon-nucleon interaction
which  dominate the resummation of loop or ladder diagrams,
respectively. However, the proper description of relaxation or damping
phenomena is found to require both types of diagrams as well as the
mixed terms simultaneously.
\end{abstract}
\newpage
\zeqn
\section{Introduction}
The interplay between single-particle (or collisional) and collective
motion is still a fundamental and fascinating problem in nuclear dynamics.
These two competing mechanisms are usually attributed to the different
effects of short- and  long-range correlations, respectively.
However, the complicated evolution of these correlations, as described by
the BBGKY density matrix hierarchy \cite{1}, does not allow to disentangle them
in a straightforward way.
Hence one usually restricts oneself to the consideration of one type
of correlations depending on the problem under investigation:
short-range correlations are taken into account by
Brueckner G-matrix approaches \cite{2}-\cite{3e} - as used e.g. in
transport equations \cite{4,5,5a} - to probe the nuclear
equation of state (EOS) or the in-medium two-body cross section,
while long-range correlations are included in extended RPA
calculations in order to describe the damping of collective
nuclear motion \cite{6,7,7a,7b,7c,21}.
These separate applications are based on the concept
that the sum of ladder diagrams should be essential at large
momentum transfer whereas one expects loop diagrams to be dominant
at small momentum transfer \cite{7d,7e}.
Our approach aims at the simultaneous consideration of both types
of correlations.
It is based on the coupled system of equations for the one-body
density matrix and the two-body correlation function \cite{8}-\cite{13}.
Due to the structure of the two-body equation the formation of
correlations proceeds simultaneously in two "orthogonal" channels:
the particle-particle ($pp$) channel which accounts for a resummation of
ladder diagrams and hence for two-body collisions, and the
particle-hole ($ph$) channel which accounts for a resummation of loop
diagrams and hence for collective or coherent motion.
Due to the discussion by Abrikosov et al. \cite{7d}
for infinite matter and by Migdal \cite{7e} for finite nuclei it is
tempting to associate short- and long-range correlations with
the particle-particle ($pp$) and the particle-hole ($ph$) channel,
respectively. This becomes obvious in the stationary limit,
where the two-body equation can be traced back to a Brueckner-like
integral equation for the G-matrix when neglecting the $ph$-channel,
and an integral equation for the polarization matrix when neglecting
the $pp$-channel \cite{9,10,Krieg}.

One goal of the present paper is to get some insight into the relative
importance of long- and short-range correlations as well as
their interference.
This interference due to a mixing of loop and ladder diagrams
has been also addressed within the framework of Green's functions
\cite{11a,11b}.
In order to test such effects in light nuclei we study the influence of
correlations on the distribution of occupation numbers, the relaxation of
single-particle excitations and the damping of collective motion depending
on the range of the nucleon-nucleon interaction and the approximations
applied on the two-body level.
Detailed information about the dynamics is obtained by
discriminating between the formation of different types of
correlations which are defined with respect to the
Hartree-Fock groundstate.

The present paper is organized as follows:
In sect. 2 we present a brief recapitulation of the equations of motion
for the one-body density matrix and the two-body correlation function as
used in this work within different limites.
For a more detailed presentation and discussion we refer
the reader to [18-20,25].
In addition to the calculations in ref. \cite{12} we, furthermore,
include the trace conserving terms in the two-body equation
 - introduced in ref. \cite{10} -
which simulate a part of three-body dynamics on the two-body level.
In order to reestablish the relation to known many-body approaches
in the stationary limit
we derive the integral equations already mentioned above for the corresponding
effective interactions in the $pp$ and $ph$ channel.

Since the problem of nonperturbative trace-conserving approaches has been
raised recently in the literature \cite{11,13} we show that the
 trace conservation is restored on a sufficiently
long time scale. Furthermore, the influence of these
trace conserving terms on the nuclear dynamics is investigated in detail
throughout the paper.

In sect. 3 we study the redistribution of the occupation numbers for a
configuration close to the groundstate due to the presence of
two-body correlations.
The relaxation of single-particle excitations is considered in
sect. 4, while sect. 5 is devoted to a detailed study of the
damping of isoscalar giant monopole resonances (ISGMR) in $^{16}O$ for
different nuclear incompressibilities. In order to present our common
findings from sections 3-5 in a unified way
we investigate the time evolution of different types of
correlations close to the groundstate in sect. 6 which will allow for
a transparent picture on the formation and effects of two-body
correlations. The conclusions are summarized in sect. 7 while
necessary numerical details are presented in the Appendices.
\zeqn
\section{General aspects of correlation dynamics}
\subsection{The equations of motion}
In the nonrelativistic limit (restricting to the two-body level)
the nuclear many-body problem can be reformulated in terms of coupled
equations of motion for the one-body density matrix $\rho_{\alpha \alpha'}$
\bea
i\frac{\partial}{\partial t} \rho_{\alpha\alpha'} =
\langle\alpha|[h(1),\rho]|\alpha'\rangle +
\sum_{\beta} \langle\alpha\beta|[v,c_2]|\alpha'\beta\rangle
\label{2.1}
\eea
and the two-body correlation function $C_{\alpha \beta \alpha' \beta'}$
\bea
i\frac{\partial}{\partial t} C_{\alpha\beta\alpha'\beta'} =
\langle\alpha\beta|[h(1)+h(2),c_2]|\alpha'\beta'\rangle
\nonumber\\ \label{2.2} \eqnoinc
+ \langle\alpha\beta|[V^=,\rho_{20}]|\alpha'\beta'\rangle \qquad\qquad
\subeqno \label{2.2a}
\\
+ \langle\alpha\beta|[V^=,c_2]|\alpha'\beta'\rangle  \qquad\qquad
\nonumber\\
\subeqno \label{2.2b}
\\
+ {\cal A}_{\alpha\beta} {\cal A}_{\alpha'\beta'} \sum_{\gamma\gamma'}
\langle\alpha\gamma'|V^\perp|\alpha'\gamma\rangle_{\cal A}
C_{\gamma\beta\gamma'\beta'}
\subeqno \label{2.2c}
\\
+ \sum_{\gamma \gamma'} \delta_{\gamma \gamma'}
({\cal P}_{\alpha\gamma} {\cal P}_{\alpha'\gamma'}
+{\cal P}_{\beta\gamma} {\cal P}_{\beta'\gamma'})  \nonumber \\
 \times \sum_{\lambda \delta}
\lbrace
\langle\alpha\beta|v|\lambda \delta \rangle
C_{\lambda \delta \gamma \alpha'\beta'\gamma'}
- C_{\alpha \beta \gamma \lambda \delta \gamma'}
\langle \lambda \delta|v|\alpha' \beta' \rangle \rbrace
\subeqno \label{2.2d}
\eea
\subeqres
which are obtained by an expansion of the corresponding
coordinate-space equations for $\rho$ and $c_2$
\cite{8} - \cite{13} within an arbitrary
complete and orthogonal single-particle basis
$|\alpha\rangle$.

In eqs.(\ref{2.1}, \ref{2.2}) we have introduced the mean-field
hamiltonian
$h(i) = t(i) + Tr_2\lbrace v(i,2) {\cal A}_{i2} \rho(22')\rbrace $,
the antisymmetrization operator ${\cal A}_{\alpha \beta} = 1 - {\cal P}_{
\alpha \beta}$ with the permutation operator ${\cal P}_{\alpha \beta}$
and the "horizontal" and "vertical" in-medium interactions,
\bea
\langle\alpha \beta |V^=|\alpha '\beta '\rangle &=& \sum_{\lambda \gamma }
Q^=_{\alpha \beta \lambda \gamma } \langle\lambda \gamma |v|\alpha '\beta
'\rangle
\label{2.3}\\
\langle\alpha \gamma '|V^\perp |\alpha '\gamma \rangle_{\cal A} &=&
\sum_{\lambda
\lambda '} Q^\perp _{\alpha \lambda \alpha '\lambda '}
\langle\lambda '\gamma '|v|\lambda \gamma \rangle_{\cal A}\;,
\label{2.4}
\eea
by combining the matrix elements of
the particle-particle ($pp$) and particle-hole ($ph$)
blocking operators
\bea
Q^\perp _{\alpha \lambda \alpha '\lambda '} &=&  \delta _{\alpha \lambda '}
\rho _{\lambda \alpha '} -  \rho _{\alpha \lambda '}
\delta _{\alpha '\lambda }
\label{2.5}\\
Q^=_{\alpha \beta \lambda '\gamma '} &=&
\delta _{\alpha \lambda '}\delta _{\beta \gamma '} -
\delta _{\alpha \lambda '}\rho _{\beta \gamma '} -
\rho _{\alpha \lambda '}\delta _{\beta \gamma '}
\label{2.6}
\eea
with the matrix elements of the two-body interaction $v$.
The two-body density matrix can be written as
\bea
\rho^2_{\alpha \beta \alpha '\beta '} =
(\rho_{20})_{\alpha\beta\alpha'\beta'} + C_{\alpha\beta\alpha'\beta'}
= {\cal A}_{\alpha \beta} \lbrace\rho_{\alpha\alpha'}\rho_{\beta\beta'}\rbrace
   + C_{\alpha\beta\alpha'\beta'}    \; ,
\label{2.7}
\eea
where $(\rho_{20})_{\alpha\beta\alpha'\beta'}$ represents the
uncorrelated two-body density matrix.

According to the physical content of the in-medium interactions
(\ref{2.3}) and (\ref{2.4})\footnote{In a single-particle basis
$|\alpha\rangle$ that diagonalizes the one-body density matrix,
one can easily verify that $Q^=$ leads to a projection
on the $pp$ and $hh$ channel while $Q^\perp$ projects on the $ph$ channel
(see e.g. \cite{12}).}  it is
obvious that (\ref{2.2a}) which accounts for $pp$ collisions in
Born approximation ({\bf BORN}) together with (\ref{2.2b}) correspond to a
nonperturbative resummation of ladder diagrams
({\bf TDGMT}\footnote{Abbreviation for Time-Dependent G-Matrix Theory
according to \cite{9,10}})
while
(\ref{2.2a}) together with (\ref{2.2c}) take into account interactions in the
$ph$-channel by resumming the loop diagrams ({\bf RPA}).
A simultaneous inclusion of these two "orthogonal" channels can be
achieved by incorporating (2.2a+b+c) in the calculations.
This approach is denoted by {\bf NQCD} (nuclear quantum correlation dynamics)
as in \cite{9,10}.
Eq. (\ref{2.2d}) comprises the coupling of two-body dynamics to the
three-body level via the three-body correlation function $c_3$.
In order to close the system of eqs. (\ref{2.1}) and (\ref{2.2}) one
usually neglects $c_3$. However, this
truncation dynamically violates the conservation of fundamental trace relations
between the one- and two-body density-matrices \cite{10,13}, i.e.
\bea
(A-1)\rho_{\alpha\alpha'} =  \sum_{\beta}
\rho^2_{\alpha\beta\alpha'\beta}
\label{2.8}
\eea
is violated in time where A denotes the number of nucleons.
 This can be directly related to the violation of
\bea
(A-2) \rho^2_{\alpha\beta\alpha'\beta'} =  \sum_{\gamma}
\rho^3_{\alpha\beta\gamma\alpha'\beta'\gamma}
\label{2.9}
\eea
at arbitrary times in case of $c_3=0$. The problem can be cured by specifying
the matrix elements of $c_3$ in eq. (\ref{2.2d}) according
to the ansatz \cite{10}
\bea
C_{\alpha\beta\gamma\alpha'\beta'\gamma'} =
  \frac{-1}{(A-1)} \sum_{\lambda}
\left(\frac{2}{A}-{\cal P}_{\alpha\lambda}-{\cal P}_{\beta\lambda}-
{\cal P}_{\alpha'\lambda}-{\cal P}_{\beta'\lambda} \right)
 \rho^2_{\gamma\lambda\gamma'\lambda}C_{\alpha\beta\alpha'\beta'}
\label{2.10}
\eea
which leads to a dynamically conserved trace relation (\ref{2.8}).
For a more detailed representation of eq. (\ref{2.2d}) with $c_3$
specified as in eq. (\ref{2.10}) we refer the reader to ref. \cite{10}.
This consistent approach on the two-body level is denoted by {\bf SCD}
(selfconsistent correlation dynamics). We recall that though the trace
relations are exactly conserved in the limit SCD, the antisymmetry
of (2.2d) with respect to an interchange of particle 1 (index $\alpha$)
and 2 (index $\beta$) is partly violated.
Thus the SCD equations should only be applied to spin-symmetric systems
where the spin degree of freedom can be traced out (see below).

In order to obtain effective interactions in the $pp$- and $ph$-channel
separately we introduce channel correlations $C^=$ and $C^\perp$,
respectively, by dropping  the corresponding terms (2.2c) or (2.2b)
in the two-body equation. In the stationary limit $C^=$ follows
\bea
\lbrace
\omega - ( \epsilon_{\alpha} + \epsilon_{\beta} )
\rbrace
C^=_{\alpha\beta\alpha'\beta'} -
\lbrace
\omega - ( \epsilon_{\alpha'} + \epsilon_{\beta'} )
\rbrace
C^=_{\alpha\beta\alpha'\beta'}
\nonumber\\ \nonumber\\
=\sum_{\lambda\gamma}
\lbrace
\langle\alpha\beta|V^=|\lambda\gamma\rangle
\rho^2_{\lambda\gamma\alpha'\beta'}
-\rho^2_{\alpha\beta\lambda\gamma}
\langle\lambda\gamma|V^{=\dagger}|\alpha'\beta'\rangle
\rbrace \;,
\label{2.11}
\eea
assuming $h_{\alpha\lambda} = \epsilon_{\alpha} \delta_{\alpha\lambda}$
and using
$i\partial /\partial t C^=_{\alpha\beta\alpha'\beta'} =
0 = (\omega - \omega) C^=_{\alpha \beta \alpha' \beta'}$ on
the $l.h.s.$ in order to allow for a separation of the
different amplitudes.
Eq. (\ref{2.11}) is fulfilled by ($\varepsilon \rightarrow 0^+$)
\be
C^=_{\alpha\beta\alpha'\beta'} =
\frac{1}{\lbrace \omega-(\epsilon_{\alpha}+\epsilon_{\beta}) + i \varepsilon
\rbrace}
\sum_{\lambda\gamma}
\langle\alpha\beta|V^=|\lambda\gamma\rangle
\rho^2_{\lambda\gamma\alpha'\beta'}
\label{2.12}
\ee
and its hermitean conjugate equation. Defining a ${\cal G}$-matrix via
\be
\langle\alpha\beta|v(\rho_{20}+c_2^=)|\alpha'\beta'\rangle\: = \:
\langle\alpha\beta|{\cal G} \rho_{20}|\alpha'\beta'\rangle
\label{2.13}
\ee
and using eq. (\ref{2.7}) we obtain for arbitrary matrix elements of  the
uncorrelated two-body density matrix $\rho_{20}$
\bea
\langle\alpha\beta|{\cal G}|\alpha'\beta'\rangle =
\langle\alpha\beta|v|\alpha'\beta'\rangle
+\qquad\qquad\qquad
\nonumber\\
\sum_{\lambda\gamma\lambda'\gamma'}
\langle\alpha\beta|v|\lambda\gamma\rangle
\frac{Q^=_{\lambda\gamma\lambda'\gamma'} }
{ \lbrace \omega - (\epsilon_{\lambda}+\epsilon_{\gamma})+ i\varepsilon \rbrace
}
\langle\lambda'\gamma'|{\cal G}|\alpha'\beta'\rangle \;\; ,
\label{2.14}
\eea
which is the ${\cal G}$-matrix equation. We note that eq. (\ref{2.14})
contains additional $hh$ scattering in contrast to conventional
Brueckner theory\footnote{In a single-particle basis that diagonalizes the
one-body density matrix
$\rho_{\alpha\alpha'}=n_\alpha \delta_{\alpha\alpha'}$, the matrix elements
of $Q^=$ are given by
$Q^=_{\lambda\gamma\lambda'\gamma'} = \delta_{\lambda\lambda'}
\delta_{\gamma\gamma'} (1 - n_\lambda - n_\gamma)
= \delta_{\lambda\lambda'} \delta_{\gamma\gamma'}
\lbrace (1-n_\lambda) (1-n_\gamma) - n_\lambda n_\gamma \rbrace $.}.

In the $ph$  channel one can proceed in a similar  way by using that the
Born term (\ref{2.2a}) can be rewritten with "vertical" instead of
"horizontal" blocking operators, i.e.
\bea
\sum_{\lambda\gamma\lambda'\gamma'} \lbrace
Q^=_{\alpha\beta\lambda'\gamma'}
\langle\lambda'\gamma'|v|\lambda\gamma\rangle
(\rho_{20})_{\lambda\gamma\alpha'\beta'}
-
(\rho_{20})_{\alpha\beta\lambda'\gamma'}
\langle\lambda'\gamma'|v|\lambda\gamma\rangle
Q^=_{\lambda\gamma\alpha'\beta'} \rbrace
\nonumber\\
=\sum_{\lambda\gamma\lambda'\gamma'}
\langle\lambda'\gamma'|v|\lambda\gamma\rangle_{\cal A}
\lbrace
Q^\perp_{\alpha\lambda\alpha'\lambda'}\: \rho_{\gamma\beta'}
( 1 - \rho_{\beta\gamma'} )
+ Q^\perp_{\beta\lambda\beta'\lambda'}\: \rho_{\alpha\gamma'}
( 1 - \rho_{\gamma\alpha'} )
\rbrace
\label{2.15}
\eea
which shows that they cannot be attributed to a specific channel alone.
Realizing that ${\cal A}_{\alpha\beta}{\cal A}_{\alpha'\beta'}
= {\cal A}_{\alpha \beta} (1+{\cal P}_{\alpha\beta}{\cal P}_{\alpha'\beta'})$
in (2.2c) and neglecting the exchange term we obtain
in the stationary limit of (2.2a+c)
\bea
\lbrace \omega - \epsilon_{\alpha} + \epsilon_{\alpha'} \rbrace
C^\perp_{\alpha\beta\alpha'\beta'}
- \lbrace \omega + \epsilon_{\beta} - \epsilon_{\beta'} \rbrace
C^\perp_{\alpha\beta\alpha'\beta'}
\nonumber\\
\nonumber\\
=\sum_{\gamma\gamma'}
\langle\alpha\gamma'|V^\perp|\alpha'\gamma\rangle_{\cal A}
\lbrace C^\perp_{\gamma\beta\gamma'\beta'}
+ \rho_{\gamma\beta'} ( 1 - \rho_{\beta\gamma'} ) \rbrace
\nonumber\\
+ \sum_{\gamma\gamma'}
\langle\beta\gamma'|V^\perp|\beta'\gamma\rangle_{\cal A}
\lbrace C^\perp_{\gamma\alpha\gamma'\alpha'}
+ \rho_{\alpha\gamma'} (1 - \rho_{\gamma\alpha'}) \rbrace
\label{2.16}
\eea
with the solution $(\varepsilon\rightarrow 0^+)$
\bea
C^\perp_{\alpha\beta\alpha'\beta'} &=&
\frac{1}{ \lbrace \omega -\epsilon_{\alpha}
+\epsilon_{\alpha'} + i \varepsilon \rbrace }
\sum_{\gamma\gamma'}
\langle\alpha\gamma'|V^\perp|\alpha'\gamma\rangle_{\cal A}
\lbrace C^\perp_{\gamma\beta\gamma'\beta'}
+ \rho_{\gamma\beta'} ( 1 - \rho_{\beta\gamma'} ) \rbrace .
\nonumber\\
\label{2.17}
\eea
Again we have added the identity
$i\partial /\partial t C^\perp_{\alpha\beta\alpha'\beta'} = 0 =
(\omega - \omega )C^\perp_{\alpha\beta\alpha'\beta'}$
on the $l.h.s$ of eq. (\ref{2.16}) in order to allow for the separation
of the two different amplitudes.
Defining a polarization matrix $\Pi$ via
\be
\sum_{\gamma \gamma'} \langle\alpha\gamma'|v|\alpha' \gamma\rangle_{\cal A}
(C^{\perp}_{\gamma \beta \gamma'\beta'} + \rho_{\gamma \beta'}
(\delta_{\beta \gamma'} - \rho_{\beta \gamma'})) =
\sum_{\gamma \gamma'}  \langle \alpha\gamma'|\Pi|\alpha' \gamma\rangle_{\cal A}
\rho_{\gamma \beta'} (\delta_{\beta \gamma'} - \rho_{\beta \gamma'})
\label{2.18}
\ee
one obtains for arbitrary matrixelements of $\rho (1-\rho)$  the equation
\bea
\langle\alpha\beta|\Pi|\alpha'\beta'\rangle_{\cal A} \:=\:
\langle\alpha\beta|v|\alpha'\beta'\rangle_{\cal A}\qquad\qquad\qquad
\nonumber\\
 +
\sum_{\gamma\gamma'\lambda\lambda'}
\langle\alpha\gamma|v|\alpha'\gamma'\rangle_{\cal A}
\frac{Q^\perp_{\gamma'\lambda'\gamma\lambda}}
 {\lbrace \omega - \epsilon_{\gamma'} + \epsilon_{\gamma} + i \varepsilon
\rbrace}
 \langle\lambda\beta|\Pi|\lambda'\beta'\rangle_{\cal A}
\label{2.19}
\eea
which leads to a nonperturbative resummation of loop diagrams.

Of course one should keep in mind that a separation of either loop or ladder
diagrams neglects the possibly large contributions of "mixed"
diagrams which reflect the mutual influence of the two "orthogonal" channels
(parquet diagrams). We will address the relative importance of these mixed
diagrams throughout the paper.
\subsection{Specification of the model}
The coupled set of eqs. (\ref{2.1}) and (\ref{2.2}) has already been
used for microscopic studies on the damping of giant
resonances at finite temperature \cite{14,15,16} and the investigation of
mass fluctuations in heavy ion collisions \cite{16a}.
This approach denoted by {\bf TDDM} (time-dependent density-matrix theory)
is characterized by a realistic single-particle basis
$\lbrace\psi_{\alpha}(t)\rbrace$ following the TDHF-like equation
\be
\left\lbrace i \frac{\partial}{\partial t} - h(\rho) \right\rbrace
 \psi_{\alpha}(t) = 0
\label{2.20}
\ee
using a Bonche-Koonin-Negele (BKN) force \cite{17} for the TDHF
hamiltonian $h(\rho)$ and a contact interaction $v({\bf r-r'}) = V_0
\delta^3({\bf r-r'})$ with
$V_0=-300$ MeV ${\rm fm}^3$ in the two-body equation.

The main motivation of the present numerical study is
the question about the relative importance of long- and short-range
correlations in the dynamics of strongly interacting fermion systems, which
implies to go beyond the limit of a $\delta$-force for the residual
interaction and to implement finite-range two-body potentials. The
tremendous computational effort requires to introduce a more simplified
basis which, however, should be close to a Hartee-Fock basis in case of
light nuclei which we want to address. In view of this
we solve the coupled equations for $\rho$ and $c_2$
within a harmonic oscillator basis, which - using transformation techniques
extensively applied in nuclear structure theory \cite{29}-\cite{33}
(Appendix C) -   for the first time allows to use a more realistic
 nucleon-nucleon interaction  on the two-body level.
By comparing results generated with a contact interaction
$ v=V_0  \delta^3({\bf  r_1 - r_2}) $ with $V_0=-220$ MeV ${\rm fm}^3$ and a
long-range Yukawa potential
\be
v_y({\bf r_1, r_2}) = \frac{V_{0 \pi}(\hbar c)}{|{\bf r_1 - r_2}|}
exp(-\frac{m_\pi}{\hbar c} |{\bf r_1 - r_2}|) +
 \frac{V_{0 \omega} (\hbar c)}{|{\bf r_1 - r_2}|}
exp(-\frac{m_\omega}{\hbar c} |{\bf r_1 - r_2}|)
\label{2.21}
\ee
consisting of a repulsive $\omega$-exchange
contribution $(V_{0 \omega} = 19.0)$ and an attractive $\pi$-exchange term
$(V_{0 \pi} = -3.4)$, a clear separation between effects generated by the
long-range and the short-range part of the potential can be achieved.

Throughout the paper we will compare our results obtained
within the harmonic oscillator basis with calculations in the TDDM
approach in order to illustrate the physical significance of
the presented numerical investigations and the basis invariance of the
physical statements.
Vice versa it should be possible to gain some insight into the
shortcomings of the present TDDM approach caused by
restricting to unrealistic $\delta$-interactions on the two-body level.

In order to reduce the computation time and to be consistent with the
SCD equations we restrict ourselves to spin-isospin
symmetric light systems which allows to average the equations
of motion (\ref{2.1}) and (\ref{2.2}) over spin and isospin according to
\bea
\rho_{\alpha\beta} &=& \frac{1}{4}
\sum_{\sigma_{\alpha} \tau_{\alpha}}
\rho_{\tilde{\alpha} \tilde{\beta}}
\qquad\qquad\qquad
\nonumber\\
C_{\alpha\beta\alpha'\beta'} &=& \frac{1}{16}
\sum_{ \sigma_{\alpha} \tau_{\alpha} }
\sum_{ \sigma_{\beta}  \tau_{\beta}  }
C_{\tilde{\alpha} \tilde{\beta} \tilde{\alpha'} \tilde{\beta'}}
\label{2.22}
\eea
within the assumptions
\bea
\rho_{\tilde{\alpha} \tilde{\beta}} &=&
\rho_{\alpha\beta}\:
\delta_{\tau_{\alpha} \tau_{\beta}}
\delta_{\sigma_{\alpha} \sigma_{\beta}}
\qquad\qquad\qquad
\nonumber\\
\langle\tilde{\alpha} \tilde{\beta} |v| \tilde{\alpha'} \tilde{\beta'} \rangle
&=& \langle\alpha\beta|v|\alpha'\beta'\rangle
\delta_{\tau_{\alpha}\tau_{\alpha'}}
\delta_{\tau_{\beta}\tau_{\beta'}}
\delta_{\sigma_{\alpha}\sigma_{\alpha'}}
\delta_{\sigma_{\beta}\sigma_{\beta'}} \;\;,
\label{2.23}
\eea
where $\tilde{\alpha}$ denotes
$\lbrace\alpha,\sigma_\alpha,\tau_\alpha\rbrace$.
The resulting equations of motion are integrated in time using a
standard Runge-Kutta method with timestepsize
$\Delta t = 0.5 \times 10^{-23}s$ adopting a Hartree-Fock initial
state of some given thermal excitation energy $E^*$ or temperature $T$, i.e.
\bea
\rho_{\alpha\alpha'}(\epsilon_{\alpha};t=0) &=& 0 \qquad (\alpha \neq \alpha')
\nonumber\\
\rho_{\alpha\alpha}(\epsilon_{\alpha};t=0) &=&
\left[ 1 + exp \left\lbrace \frac{ \epsilon_{\alpha}-\epsilon_{F}}{T}
\right\rbrace
\right]^{-1}
\nonumber\\
C_{\alpha\beta\alpha'\beta'}(t=0) &=& 0 \;.
\label{2.24}
\eea
After a sufficiently long time of integration - typically around
$20 \times 10^{-23}$s - one numerically generates a correlated nucleus.
However, the question arises how much excitation energy in addition
to the simulated thermal excitation energy is imposed to the system due
to the initial conditions (2.24) which correspond to an uncorrelated
system at $t = 0$.

Since the question concerning the "true" groundstate of the
nucleus is far beyond the scope of this paper - and a matter of separate
analysis - we estimate the imposed excitation energy by analyzing
 the fluctuation properties of one-body operators, in particular the collective
isoscalar quadrupole mode of $^{40}Ca$ within the TDDM approach.
In this respect we boost the correlated $^{40}Ca$ nucleus
 at $t = 20 \times 10^{-23}$s
by applying appropriate phase factors proportional to a strength
factor $\alpha$ (for more details the reader is referred to \cite{15,16}).
The collective excitation energy gained by the system in this way
 is about $21$ MeV.
After a sufficiently long time the dispersion of the quadrupole operator
(\ref{a2}) oscillates around a fixed mean value which is slightly above
the zero-point motion value (cf. \cite{16}). However, the corresponding
asymptotic fluctuations in the quadrupole velocity
$\sigma^2_{\dot{Q}} (\infty)$ (\ref{a1}, \ref{a3}), multiplied by the
collective
mass parameter $M_Q$ (\ref{a5}), are found to follow the equipartition
theorem at higher $T$ \cite{16}. The correlation between the 'physical'
 temperature and our initialization
parameter $T$  is displayed in fig. 1.
The deviation of the calculated points for low initial
'temperature' $T$ from the straight line predicted by
the equipartition theorem (\ref{a4}) now offers a direct measure
for the energy difference to the "true" groundstate.
Hence an initialization with $T=0$ MeV corresponds to a $^{40}Ca$-nucleus
with a realistic temperature of 0.7 MeV, i.e. a thermal excitation
 energy $E^* \approx 40/8 \cdot 0.7^2 \;{\rm MeV} < 2.5 \;{\rm MeV}$ above the
"true"
groundstate. Similar values are also obtained for the case of $^{16}O$ such
that any discussion on temperature dependent quantities (see below) should
be valid in the range $ 1\; {\rm MeV} < T < 5\; {\rm MeV}$.

We now continue with general properties of our dynamical model that have to
be clarified before studying the dynamical responses.
The trace-conserving properties of the terms (\ref{2.2d}) with $c_3$
specified according to eq. (\ref{2.10}) so far have been
tested numerically in ref.\cite{10} by a one-dimensional
computation in phase-space.
To ensure that this also works within our present 3-dim. model
 we present in fig. 2
the numerical results for the quantity $H_{diff}(t)$ defined
in ref. \cite{10}, i.e.
\bea
H_{diff}(t) =
 \sum_{\alpha\alpha'}
\left| \rho_{\alpha\alpha'}(t) - {\tilde \rho}_{\alpha\alpha'}(t) \right| /
\sum_{\alpha\alpha'} \left| \rho_{\alpha\alpha'}(t)
\right|\;\; ,
\label{2.25}
\eea
with ${\tilde \rho}_{\alpha\alpha'}(t)$ given by
\bea
{\tilde \rho}_{\alpha\alpha'}(t) =
\frac{4}{(A-1)} \sum_{\gamma}
\left\lbrace
\rho_{\alpha\alpha'}(t)\rho_{\gamma\gamma}(t)
-\frac{\rho_{\alpha\gamma}(t)\rho_{\gamma\alpha'}(t)}{4}
+ C_{\alpha\gamma\alpha'\gamma}(t)
\right\rbrace.
\label{2.26}
\eea
This quantity - which should be zero according to the trace relations -
is calculated for $A=16$ using the contact interaction.
A comparison with the results in
ref. \cite{10} shows that in our model the trace conservation in the limit
SCD is improved numerically by about two orders of magnitude and
 - even more important -
is tested on a much longer time scale (see insertion in fig. 2).
The comparison between the results for $H_{diff}(t)$ obtained with
different timestepsizes $\Delta t = 0.50 \times 10^{-23} s$ or $0.40 \times
10^{-23} s$ clearly indicates that numerical
uncertainties are responsible for the remaining
violation of the trace relation in the SCD limit, which obviously
has to be the case since the dynamical conservation of the trace
relations in this limit can be shown analytically \cite{10}.
 As in \cite{10} the
other limites, i.e. BORN, TDGMT and NQCD violate the trace relations quite
sincerely on the level of a few times $10^{-2}$.
As mentioned before, we will investigate the influence of
the trace conserving terms on the dynamical response of the system throughout
the paper.

\zeqn
\section{Redistribution of occupation numbers}
A well known indication for the presence of correlations is the
depletion of hole states and the corresponding population of
particle states with respect to the Hartree-Fock distribution.
The corresponding observable is the occupation
probability $n_\alpha=\rho_{\alpha\alpha}$ which becomes time
dependent due to the coupling to the two-body dynamics. In order to examine
the redistribution of the occupation numbers
we initialize a model $^{16}O$ nucleus at $T=0$ MeV and propagate
the system in time to generate
a correlated system in quasi-equilibrium
which is characterized by small oscillations of the occupation probabilities
around fixed mean values.
By averaging the $n_\alpha$'s in time between $20$ and $120\times 10^{-23}$s
and applying a least-square fit using a fermi-distribution as a
trial function one obtains the different curves displayed in fig. 3
depending on the approximations imposed for the equations of motion.
In addition to the calculations within the harmonic oscillator basis the upper
two plots a) and b) show numerical results within the TDDM approach
for $^{16}O$ and $^{40}Ca$ in comparison to the
experimental values for the occupation
of the valence and the first "empty" orbits as obtained from
$(e,e'p)$-reactions \cite{18}.
Due to the uncertainty of the experimentally determined
Fermi surface in fig. 3 the energy levels have been taken from the
TDDM calculations. We note again that the initialization with $T = 0$ MeV does
not exactly correspond to the groundstate of the nucleus but to a configuration
with a total excitation energy $E^* < 2.5$ MeV.

The experimental verification of a large depletion of hole states
due to correlations between the nucleons is an
indication for the existence of large groundstate correlations.
Experimental investigations from Khan et al. \cite{19} and
Eckle et al. \cite{20} for polarized deuteron scattering
on $^{34}S$ and $^{40}Ca$ targets lead to similar results.
The Fermi distributions in fig. 3a (dashed and dotted lines)
indicate that
our model $^{16}O$-nucleus has a reasonable hole and particle strength
compared with the experimental and TDDM results.
Nevertheless one should keep in mind that our results should not
be literally compared with the actual $^{16}O$-nucleus since the
possible important contribution of four-nucleon alpha-like
correlations is absent in the theory.
The negligible difference between the curves obtained for the full
theory (NQCD) with
$\delta$- (dashed line) and Yukawa interactions (dotted line)
is due to the fact that the
coupling constants are adjusted in such a way that the correlation
energy is similar for both interactions.

The lower two plots of fig. 3 c) and d) allow to extract information about
the sensitivity of the particle (hole) redistribution
 on the range of the interaction and the approximation applied with respect
to the Hartree-Fock configuration\footnote{Note that the Hartree-Fock limit
already yields a diffuse Fermi surface within the oscillator
basis adopted.}.
Obviously a reasonable hole strength can only be achieved by
applying the complete theory (NQCD) including ladder and loop diagrams
as well as mixed diagrams up to infinite order, which guarantees the
inclusion of long-range and short-range correlations, respectively.
Comparing the distributions obtained by using the
$\delta$-force and the Yukawa-potential we observe a remarkable
difference regarding the relative importance of the
different channels:
with the short-range $\delta$-force most of the depletion is
produced by TDGMT while RPA remains relatively unimportant,
whereas with the long-range Yukawa potential the situation
is just opposite since the limit RPA now generates almost the entire
redistribution.
This indicates that loop diagrams are governed by the long-range
part of the interaction while ladder diagrams are controlled by
the short-range part.
In other words:
short-range correlations are those which essentially evolve in the
$pp$-channel while long-range correlations prefer the $ph$-channel.
\zeqn
\section{Relaxation of single-particle excitations}
Besides the investigation of nuclear properties close to the  groundstate
under the influence of correlations it is interesting to study the
competition between $C^=$ and $C^\perp$ in relaxation phenomena.
Therefore we initialize a model $^{16}O$-nucleus without thermal excitation
(T = 0 MeV)
and propagate the system until $t = 50 \times 10^{-23}$s in order to allow
for the formation of a correlated state.
Then the system is excited by populating an
unoccupied basis state about 13 MeV above the Fermi surface.
Due to spin-isospin degeneracy such a change increases the number of
nucleons by 4.
After this manipulation the resulting $^{20}Ne$-like nucleus is
propagated in time until $t=200\times 10^{-23}$s in order to approach
to equilibrium again.
Fig. 4 shows the time evolution of the respective occupation number
after excitation
for different approximations in case of the $\delta$-interaction.
Since the system attains an approximate equilibrium configuration
one sees a depletion of the previously populated level with time towards
its equilibrium value (indicated by the dotted line). As can be seen already
from the time dependent signals, there is only a
negligible modification of the time evolution
due to the trace conserving terms in the SCD approach as compared to NQCD which
is in agreement with the results of ref. \cite{10}.

In order to obtain quantitative values for the "damping width"
we determine the average occupation $\bar{n}_{\lambda}$
of the previously populated level $\lambda$ after equilibration.
A least-square fit with the trial function
\bea
n_{\lambda}(t) = \bar{n}_{\lambda} +
(1 - \bar{n}_{\lambda}) exp(-\gamma t / 2\hbar)
\label{4.1}
\eea
allows to extract the damping width $\gamma$ which is displayed in
fig. 5 for different approximations and interactions.
In the case of the $\delta$-force we obtain a maximum damping for the
NQCD calculation while a restriction to collisions in the $pp$
(and $hh$) or $ph$ channel leads to a longer equilibration time.
By applying a Yukawa interaction we get the surprising result
that in the RPA approach the width $\gamma$ is about a factor
 of $4$ larger than
in the TDGMT and BORN approximations and still considerably larger than
the value obtained in the NQCD calculation.
These results confirm that the long-range Yukawa interaction controls
the evolution in the $ph$ channel, while the $\delta$-force favors
the $pp$ channel. Realizing that the Yukawa potential is more
realistic than a contact interaction we may conclude
 that damping phenomena are mainly governed by
loop diagrams in the RPA and mixed diagrams in the NQCD approach.
A comparison of the damping width in the NQCD and SCD calculations
confirms that the negligible sensitivity of the dynamics to
the trace conserving terms is a general feature irrespective of
the range of the interaction. We note that in case of a smaller excitation
above the Fermi surface the width $\gamma$ becomes smaller in line with
physical expectations.

\zeqn
\section{Response of collective monopole oscillations}
In addition to the damping of single-particle excitations the decay of
collective nuclear excitations offers a unique possibility to study
coherent nuclear motion in competition with incoherent
decay mechanisms. Of special interest are isoscalar modes because
their relaxation is governed by two-body correlations
while the damping of isovector modes is essentially dominated by
mean-field dynamics \cite{15}.
By investigating the ISGMR using a Yukawa potential consisting
of  density-dependent $\pi$- and $\omega$-exchange terms
it is possible to study the damping width of the resonance
as a function of the nuclear incompressibility $K$.
\subsection{The equation of state}
Since we restrict the single-particle basis in the following
to 10 different oscillator
states it is necessary to determine reasonable coupling constants
for the $\pi$- and $\omega$- part of the interaction.
By requiring
\bea
\frac{E}{A} \left(\rho_0, b_0 \right) &=& -8\: {\rm MeV} \qquad\qquad
\label{5.1} \\
\nonumber\\
\left.\frac{\partial}{\partial \rho} \left(\frac{E}{A} \right)
\right|_{\rho_0,b_0} &=& 0
\label{5.2} \\
\nonumber\\
b_0^{^{16}O} = \sqrt{ \frac{\hbar}{m\omega_0} } &\approx&
  1.01 A^{1/6} \approx  1.6\: {\rm fm} \;,
\label{5.3}
\eea
where $b_0$ denotes the oscillator parameter for normal nuclear density
$\rho_0$, we obtain the values shown in table \ref{coupl}, where a linear
density dependence of the meson masses according to
\bea
m_{\pi,\omega} (\rho) = m_{\pi,\omega} (\rho_0, b_0)
\left[ 1 + \alpha_{DD} \left( 1- \frac{\rho}{\rho_0} \right) \right]
\label{5.4}
\eea
with $m_\pi (\rho_0,b_0) = 138$ MeV and
     $m_\omega (\rho_0,b_0)=783$ MeV
was assumed.
\begin{table}
\begin{center}
\begin{tabular}
      {||r|r|r|r||} \hline
        $\alpha_{DD}$
      & $V_{0\pi}$
      & $V_{0\omega}$
      & $ K [MeV]$
                   \\ \hline \hline
 0.00   & -5.1580   & 49.9002   & 91.30  \\
 0.05   & -5.3192   & 52.8917   & 115.13 \\
 0.10   & -5.4432   & 55.1928   & 144.26 \\
 0.15   & -5.5415   & 57.0179   & 178.21 \\
 0.20   & -5.6214   & 58.5007   & 217.43 \\
 0.25   & -5.6877   & 59.7294   & 261.45 \\
 0.30   & -5.7434   & 60.7640   & 310.25 \\
                    \hline
\end{tabular}
\end{center}
\caption{Coupling constants for different density dependences
         of the meson masses and the corresponding incompressibility $K$.}
\label{coupl}
\end{table}
By calculating the total energy of the investigated model $^{16}O$-nucleus
for different values of the oscillator parameter $b$,
which is related to the density by
$\rho /\rho_0 = b_0^3 / b^3 $, it is possible to determine the
incompressibility $K$:
\bea
K = \left. k_F^2 \:\frac{  \partial^2 (E/A)}{\partial k_F^2} \right|
      _{ \rho = \rho_0 }   =
    \left.9 \: \rho^2\: \frac{ \partial^2 (E/A)}{\partial \rho^2} \right|
      _{ \rho = \rho_0 } =
     \left.b^2 \: \frac{ \partial^2 (E/A)}{\partial b^2} \right|
     _{ b = b_0 }.
\label{5.5}
\eea
The calculated values are additionally shown in tab. \ref{coupl}
while the related equations of state (EOS) are displayed in fig. 6.
We observe a close relation between the parameter $\alpha_{DD}$
and the corresponding EOS in such a way that an increasing value of
$\alpha_{DD}$ leads to an EOS with increasing incompressibility $K$.
The functional dependence is displayed in the insertion in fig. 6
in addition to a related incompressibility of infinite
nuclear matter $K_\infty \approx 3/2 K$  \cite{21}
where surface effects have been taken out. We note that commonly accepted
values of $K_\infty$ between 220 and 320 MeV \cite{21}-\cite{24}
 correspond to a physical parameter range of
$0.1 \leq \alpha_{DD} \leq 0.2$ (dotted lines in the insertion in
fig. 6).

In order to understand the close relation between $\alpha_{DD}$
and $K$ fig. 7 shows the adopted nucleon-nucleon interaction
for the densities $0.5 \rho_0$ (solid line), $\rho_0$ (dashed line),
$2 \rho_0$ (dotted line) and $\alpha_{DD}=0.3$.
While the masses of the exchange mesons decrease with increasing density
the range of the potential increases with density.
Hence high densities lead to a nucleon-nucleon potential with its
 minimum located at
large values of $r$ while in case of low densities the range of the
interaction becomes comparably small.
In the insertion in fig.7 we have displayed the energetically lowest
oscillator wavefunction $\Phi_0$ for the same three values of $\rho$.
Of course, an increasing density yields a more localized wavefunction
in coordinate space.
This opposite behaviour of the wavefunction and the interaction as
a function of the density $\rho$ becomes more important for
large values of $\alpha_{DD}$.
Since the two-body matrix elements $\langle\alpha\beta|v|\alpha'\beta'\rangle$
are given by the overlap between the potential and the
wavefunctions it is plausible that large values of $\alpha_{DD}$
correspond to a hard EOS due to an increasing overlap of the wavefunction
with the repulsive part of the potential.
For low densities similar considerations can be applied.
Due to the unique relation between $\alpha_{DD}$ and $K$ it is
possible to connect the question concerning the influence of the
density dependence of the meson masses on
the ISGMR width directly with the influence of the corresponding
incompressibility $K$.

In order to investigate the ISGMR within our model approach it is convenient
to replace the static oscillator basis by a time dependent one.
Such a change implies to modify the coupled equations for
$\rho$ (\ref{2.1}) and $c_2$ (\ref{2.2}) by introducing a collective
term in the hamiltonian of the system (cf. Appendix B).
Furthermore, one now additionally has to determine
the time evolution of the oscillator basis, i.e. $ b(t)$,
 which is obtained by
requiring the conservation of the total energy. For further details the
reader is referred to Appendix B where we present the coupled equations
of motion for the oscillator parameter $b(t)$ and its time derivative
$\dot{b}(t)$.  The excitation of the
ISGMR in our model approach then corresponds to a specification of
$\dot{b}(t_0)$ which via (\ref{B.8}) is directly related to a collective
energy $E_{coll}(t_0)$.

According to fig. 6 we expect a close relation between the incompressibility
$K$ (or the parameter $\alpha_{DD}$) of the EOS and
the amplitude of the monopole oscillation.
Furthermore it is plausible to assume that the amplitude, which
is closely related to the maximum and minimum densities, modifies the
damping width of the ISGMR. Hence we investigate the density oscillation
of our model $^{16}O$-nucleus for $15$ and $30$ MeV collective excitation
energy
and two different values of $\alpha_{DD}$ or $K$.
The results are plotted in fig. 8 within the TDHF and NQCD approximations.
A first observation shows that the ISGMR in the TDHF limit
is mainly an undamped density oscillation while an inclusion of
two-body correlations in the NQCD approach leads to a significant
damping of the mode. In addition one observes a remarkable but well
known relation between the incompressibility of the EOS
and the resulting amplitude and frequency of the oscillation \cite{21}.
A comparison of the density vibrations with 15 and 30 MeV  collective
energies clearly indicates that the ISGMR is not very useful
to study nuclei in high density regimes since at higher collective
 energies the system spends a longer
period of time in the low density regime.
This discrepancy between the low- and the high-density regime
is a result of the density-dependent mass parameter
$M_{Q_0} = m/(4 \langle r^2 \rangle )$ (\ref{B.7}) which becomes
larger with increasing density. Because of
$E_{{\cal C}oll}= 1/2 M_{Q_0} {\dot Q_0}^2$ (\ref{B.5}) this leads to a
decrease of the collective velocity ${\dot Q_0}(t)$ with increasing density
which dynamically forces the system to stay at lower densities
for a longer time.

Apart from the qualitative features discussed in fig. 8 we have to ensure
for the further analysis that the collective nuclear response does not
depend on the actual time at which the collective excitation
is applied to the system.
Therefore we excite our model $^{16}O$-nucleus (initialized
with $T=1$ MeV at $t = 0$)  at four different times
($t_0 = 18$, $23$, $28$, $33\times 10^{-23}$s).
Fig. 9 shows the resulting time evolution of the monopole moment
\be
Q_0(t) = \langle r^2 \rangle - \langle r^2 \rangle_0
\label{5.5a}
\ee
for $15$ and $30$ MeV collective excitation energy.
In order to simplify the comparison the different time scales are
shifted to a common origin at $t_0 = 0$. Apart from differences in the
time evolution on a longer time scale we observe that
within the first four oscillations there is nearly no dependence
of $Q_0 (t)$ on the excitation time. Thus
these investigations indicate that the physical response of the
strongly interacting system - especially the collective frequency and
the damping rate - does not depend on the numerical initialization.
\subsection{The damping width in different approximations}
Since nuclear collective motion and its damping is a many-body problem
of high complexity beyond the mean-field level we have to consider
two-body correlations explicitly.
In order to study the importance of different types of two-body
correlations - as naturally arising from the two "orthogonal" channels -
we investigate the damping of the ISGMR within the different approximations
as before. The actual numerical realization proceeds as follows:
we initialize our model $^{16}O$-nucleus with $T=1$ MeV
temperature and propagate the system in time within the different limites
(TDHF, BORN, TDGMT, RPA, NQCD, SCD) in order to generate a "correlated
nucleus".
Then the system is excited by specifying ${\dot b}(t_0)$
in such a way that the system acquires a collective energy of about
$30$ MeV. In order to have a "reasonable" value for the
incompressibility of the EOS we used $\alpha_{DD}=0.2$ which
corresponds to $K=217.43$ MeV. The resulting time evolution of $Q_0 (t)$
is displayed in fig. 10 for the different approaches.
Clearly we observe no damping in the TDHF limit
while an inclusion of $pp$ (and $hh$) interactions in the Born and
TDGMT limits leads to a slightly damped signal.
A remarkable damping, however, is achieved in the $ph$-channel (RPA) while
an inclusion of all types of diagrams in the NQCD and SCD approach
only leads to a moderate enhancement of the damping rate.
Again we stress the negligible difference between the results obtained
in the SCD and the NQCD approach. This is in agreement with
our former observation
concerning the relaxation of single-particle excitations.

In order to get quantitative values for the damping width we
have carried out least square fits under the assumption that the
signal can be described by an exponentially damped oscillator, i.e.
\be
Q_0(t) \approx Q_0(t_0) exp(- \frac{\gamma t}{2 \hbar}) .
\label{5.6a}
\ee
The results obtained in this way are $\gamma  \approx 2.2, 1.3, 4.4, 6.2$
and $6.3$ MeV in case of BORN, TDGMT, RPA, NQCD and SCD, respectively.
Again one observes that only when nonperturbatively including
$ph$ interactions, a sizeable damping of the ISGMR is obtained.
Obviously a correct description of the damping of collective
isoscalar modes can only be achieved by taking into account the full
theory (NQCD or SCD) which means the nonperturbative inclusion of loop, ladder
and mixed diagrams.
\subsection{Temperature dependence of the damping width}
In this section we study the influence of thermal excitations on the damping
of the ISGMR for different values of the parameter $\alpha_{DD}$
or the incompressibility $K$. The motivation for this investigation arises
from recent experimental data which clearly show a saturation of the
GDR (giant dipole resonance) width above $1$ MeV thermal excitation
\cite{25,26,27}. These results, which are in contrast to naive expectations,
are supported by microscopic calculations e.g. by Bortignon et al.
\cite{28} which indicate only a weak dependence of the GDR spreading width
on temperature.
More recent calculations by De Blasio et al. \cite{15} within the
TDDM approach confirm these results beyond $1$ MeV temperature for
$^{40}Ca$ showing a remarkable insensitivity of the
spreading width. By investigating different possible approximations on
the two-body level De Blasio et al. have shown that the GDR width
is mainly governed by mean-field effects while the ISGQR
(isoscalar giant quadrupole resonance) width is dominated by
residual $ph$ correlations, which is in line with our results obtained
in the previous section for the ISGMR.
Besides the different nature of the damping mechanism in isoscalar
and isovector modes it is surprising that in the TDDM approach the
spreading width even in the isoscalar case remains insensitive to the
thermal excitation energy.
Hence in case of isoscalar modes one can assume a close relationship
between the temperature independence of the width and the influence
of the thermal excitations on the correlations especially in the
$ph$ channel.

 Since there are no experimental results available concerning
the temperature dependence of isoscalar modes it seems to be
interesting and necessary to study the influence of the temperature
on the ISGMR width in our model.
As our results in the previous section clearly indicate
that a reasonable description of the damping width can only be achieved
by nonperturbatively incorporating loop and ladder diagrams,
we perform the following calculations within the NQCD approach
neglecting the trace conserving terms.
At this point it is important to note that the damping obtained
in our studies is due to the two-body spreading width, whereas continuum
effects (e.g. particle evaporation) are not included.
These effects become important for temperatures above $3-4$ MeV
and should be included in an actual comparison to
experimental data which, however, are not available yet.

In order to investigate the influence of the parameter
$\alpha_{DD}$ or the incompressibility $K$ as well, we carry out the
following calculations for $\alpha_{DD}=0.00$, $0.10$, $0.20$ and $0.30$.
We initialize our model $^{16}O$-nucleus with $T=0$, $1$, $2$, $3$ and
$4$ MeV temperature and propagate the system until $t_0=18 \times 10^{-23}$s.
Then we excite the ISGMR mode of the equilibrated fully correlated nucleus
by imposing $15$ MeV collective energy.
The damping widths which are obtained according to eq. (\ref{5.6a})
are displayed in fig. 11 where the dashed curves represent fits through
the evaluated data points.
One observes a quite strong influence of $\alpha_{DD}$ (or $K$) on
the temperature dependence of the damping width.
In case of $\alpha_{DD}=0.00$ there is a strong monotoneous increase
of the width as a function of temperature from $\gamma\approx 1$ MeV
at $T=0$ MeV to about $\gamma \approx 3.5$ MeV at $T=4$ MeV while for
$\alpha_{DD}=0.10$ to $0.30$ one sees an approximate
independence of the damping width on temperature.
These results can immediately be related to the different
incompressibilities of the EOS for different values of $\alpha_{DD}$.
Since for $\alpha_{DD}=0.00$ one obtains a very soft EOS with
$K=91.3$ MeV one favors the excitation of surface fluctuations.
The coupling of the ISGMR to these modes leads to the observed
increase of the width with temperature.
This result is in contradiction to the experimental investigations
\cite{25,26,27} which show a remarkable insensitivity of the damping
width beyond $T=1$ MeV temperature.
Hence one can rule out $\alpha_{DD}=0.00$ from the
physical significant parameter range.
Finally we note that the temperature independence of the ISGMR width
in the reasonable parameter range is in line with results obtained
in the TDDM code for $^{16}O $ and $^{40}Ca$.
\subsection{Monopole-response on the compressibility}
In this section we aim at studying the influence of the density-dependent
meson masses given by the parameter $\alpha_{DD}$ or equivalently (in our
model) the incompressibility $K$ on the ISGMR width.
According to our results in section 5.2 we carry out the calculations
in the NQCD approach.
First we initialize our model $^{16}O$-nucleus at $T=0$ MeV and excite
the system at $t_0=18\times 10^{-23}$s in such
a way that the system acquires $15$ MeV collective energy.
In order to get sufficient information concerning the
long time behaviour of the monopole moment $Q_0(t)$ we propagate
the excited system until $t_f=180 \times 10^{-23}$s.
The eigenfrequencies of the system can be obtained by calculating
the fourier transform of $Q_0(t)$ according to
\be
S_0 (E) = \int^{t_f}_{t_0}  \sin(E t/\hbar) \: Q_0 (t) \: dt\; ,
\label{5.8}
\ee
where $t_0$ denotes the moment of the excitation while $t_f$
is the final time of the calculation.
By using the uncertainty relation between energy and time
it is possible to get a simple estimate concerning the quality
of the transformed signal:
\be
\Delta E \approx \frac{\hbar}{t_f-t_0} \approx 0.5\; {\rm MeV}.
\label{5.9}
\ee
In fig. 12 we have displayed $Q_0(t)$ and $S_0(E)$ for the
different values of $\alpha_{DD}$ or $K$ as specified in tab. \ref{coupl}.
The strong increase of the oscillation frequency with
increasing values of $\alpha_{DD}$ is approximately given
by a $\sqrt{K}$ dependence \cite{21}. Furthermore,
one observes a decreasing amplitude of the mode with increasing
$\alpha_{DD}$. Due to the relation between $\alpha_{DD}$ and the
incompressibility $K$ these results match with simple considerations.
More interesting is the observed ISGMR width as a function of the parameter
$\alpha_{DD}$: For $\alpha_{DD}=0.00$ the width remains small, which
yields a single sharp peak in the corresponding fourier spectrum, while
in case of $\alpha_{DD}=0.10$ and $0.15$ the width becomes larger, which
leads to a few peaks in the fourier spectrum.
The damping width approaches a maximum for about $\alpha_{DD}\approx0.20$
and decreases for higher values of $\alpha_{DD}$.

The fourier transformed signal shows the mechanism responsible for damping.
Obviously the damping is generated by a superposition of different modes
which are characterized by small differences in the frequency.
This allows to explain the observed change of the ISGMR width with
the parameter $\alpha_{DD}$ or incompressibility $K$ by a  varying
coupling of the collective mode  to
different degrees of freedom of the system.
This coupling changes with increasing compressibility and leads to
the observed results.

In order to present our results in a more quantitative way we
have determined a damping width in the usual way.
The results are displayed in fig. 13 for $15$ MeV and
additionally for $30$ MeV collective energy.
The dashed curves represent fits through the calculated values.
The physical parameter range is indicated by the dotted lines.
Again one sees the described increase
of the ISGMR width $\gamma$ up to about $\alpha_{DD} \approx 0.2$ and
the decrease for larger values of $\alpha_{DD}$.
A comparison between the results obtained with $15$ and $30$ MeV
collective energy shows that this general behaviour of $\gamma$ with
$\alpha_{DD}$ holds in both cases. But in case of $30$ MeV excitation
energy the parameter range with comparably large values of
$\gamma$ becomes larger. In fact we note that the strong
dependence of the damping width on $\alpha_{DD}$ becomes
less important if we restrict ourself to physically reasonable values
for the compressibility $K$.
Finally we would like to stress the fact that a comparison with TDHF
calculations shows that the observed behaviour is due to the explicit
inclusion of the two-body dynamics, i.e. to the dynamically
evolving correlations.

\zeqn
\section{Decomposition of two-body correlations}
The similarity of the results obtained by investigating the
time average of the occupation
probabilities and the damping of single-particle as well as collective
excitations suggests that the dynamical response is driven by the same
particular correlations in all cases. In this section we thus extract more
 detailed information about
the specific type of the correlations and their sensitivity to different
parts of the interaction.

In order to facilitate a "microscopic" understanding of the results
presented in the previous sections we study the formation of
correlations under different aspects. We initialize a model $^{16}O$-nucleus
at $T=0$ MeV  taking into account
13 oscillator states and propagate the system within different
approximations for the two-body equation. We use
a $\delta$-force as well as a Yukawa interaction
in order to test the role of the range of the nucleon-nucleon potential
and hence to establish the connection with short- and long-range correlations.

Detailed information about the generation of correlations is
obtained by following the time evolution of the different types
of correlations discriminated with respect to their
single-particle labels:
\bea
C_{pppp} &=& 16 \sum_{\alpha\beta\alpha'\beta'}
( 1-n_{\alpha} ) ( 1-n_{\beta} ) \:
C_{\alpha\beta\alpha'\beta'} \:
( 1-n_{\alpha'} ) (1-n_{\beta'} ) \qquad \\
\label{cpppp}
C_{hhhh} &=& 16 \sum_{\alpha\beta\alpha'\beta'}
n_{\alpha}n_{\beta}\:
C_{\alpha\beta\alpha'\beta'}\:
n_{\alpha'}n_{\beta'}  \\
\label{chhhh}
C_{pphh} &=& 16 \sum_{\alpha\beta\alpha'\beta'}
( 1-n_{\alpha} ) ( 1-n_{\beta} )\:
C_{\alpha\beta\alpha'\beta'}\:
n_{\alpha'}n_{\beta'}\\
\label{cpphh}
C_{phph} &=& 16 \sum_{\alpha\beta\alpha'\beta'}
( 1-n_{\alpha} ) n_{\beta} \:
C_{\alpha\beta\alpha'\beta'} \:
( 1-n_{\alpha'} ) n_{\beta'} \\
\label{cphph}
C_{phhh} &=& 16 \sum_{\alpha\beta\alpha'\beta'}
( 1-n_{\alpha} ) n_{\beta}  \:
C_{\alpha\beta\alpha'\beta'}\:
n_{\alpha'} n_{\beta'} \\
\label{cphhh}
C_{hppp} &=& 16 \sum_{\alpha\beta\alpha'\beta'}
n_{\alpha} ( 1-n_{\beta} ) \:
C_{\alpha\beta\alpha'\beta'} \:
( 1-n_{\alpha'}) (1-n_{\beta'} ),
\label{chppp}
\eea
which are defined with respect to the Hartree-Fock groundstate.
According to their definition $C_{pppp}$ and $C_{hhhh}$ describe
the correlations of "particles" above and "holes" below the fermi surface
while $C_{phph}$ describes $ph$ correlations.
The "$2p-2h$-amplitude" $C_{pphh}$ is related to spontaneous creation of
$2p-2h$ states, and the doorway correlations $C_{phhh}$, $C_{hppp}$
connect $1h$- and $1p$-states with $2h-1p$ and $2p-1h$ states, respectively.

In fig. 14 the respective time evolution is shown for the $\delta$-interaction
while the results obtained with the Yukawa potential are displayed
in fig. 15.
Irrespective of the applied approach and the range of the potential
the formation of correlations starts with a rapid increase of
$C_{pphh}$.
This leads to a diffuse Fermi surface and thus opens phase space for the
other correlations to become operative.
The $C_{pppp}$ component is mainly generated by the ladder diagrams
in case of the $\delta$-force, but contrary to naive assumptions there
is a large contribution from mixed diagrams.
For the Yukawa interaction one additionally has strong contributions
of the $ph$-channel to the $C_{pppp}$-correlations which shows that
simple considerations with respect to the Hartree-Fock groundstate are not
always applicable when strong correlations are involved.
A common feature is that the $C_{hhhh}$-component remains small in all cases.
The $ph$-correlations $C_{phph}$ show a typical behaviour.
Irrespective of the range of the potential they become significant in the
RPA approach, i.e. they are produced exclusively in the $ph$-channel.
The $C_{ppph}$ correlations should be especially sensitive to
mixed diagrams. This is demonstrated in case of the
Yukawa interaction where one gets no significant contributions in
the $pp$-channel while inclusion of loop and mixed
diagrams leads to a strong enhancement.

In order to get further insight into the relative importance of
the different types of two-body correlations, they are
averaged in time between $20$ and $120 \times 10^{-23}$s and
normalized with respect to the sum over all expansion coefficients
$C_{\alpha\beta\alpha'\beta'}$.
The resulting values are displayed - for both interactions -
in fig. 16 in terms of histograms.
Most important are the $pp$-correlations $C_{pppp}$
and the doorway correlations $C_{phhh} + C_{ppph}$.
Again by comparing the results generated by the two different
potentials we find that the $\delta$-force overestimates
the $pp$-channel. In case of the $2p-2h$ correlations, which
are important for the damping of isoscalar giant resonances, the
$\delta$-force leads to higher contributions than the Yukawa potential.
However, by looking at the origin of the $2p-2h$ correlations in
figs. 14 and 15 it becomes obvious that in case of the contact interaction most
of the $2p-2h$ correlations are generated already in the Born limit while
 in case of the Yukawa potential the $2p-2h$ parts additionally are
built up by loop diagrams.

In regard of our results in Section 5.3 and
the numerical studies by De Blasio et al. \cite{15},
which show no significant temperature dependence of the
damping width for isoscalar modes between
$T=0$ and $4$ MeV thermal excitation energy, it is furthermore interesting
to study the influence of the temperature on the relative importance
of the two-body correlations.
Therefore we initialize an appropriate Fermi distribution with T = 2 and
4 MeV temperature
according to eq. (\ref{2.23}) and determine the time-average of the
different types of two-body correlations.
The results are displayed in fig. 17 for the Yukawa potential in
comparison with the results obtained for $T=0$ MeV.
We get the surprising result that besides a small increase of
$C_{phph}$ with temperature the relative weight of the different
contributions remains unchanged in the range from $0$ to $4$ MeV
thermal excitation. Since the correlations altogether become less important
with increasing temperature it seems plausible to assume that the
damping of giant resonances is mainly influenced by the relative importance
of the different types of correlations and not by their absolute value.

To obtain a measure for the strength of all types of
correlations, the sum over all
matrix elements $C_{\alpha\beta\alpha'\beta'}$ is averaged over
a time period of $100\times 10^{-23}$s and normalized with respect to
the full theory. The results for the different approximations are presented
in fig. 18 for the case $T=0$ MeV in histogram representation.
By simply calculating the difference between the applied approximations
we can approximately  separate higher ladder diagram contributions
$[V^=,c_2]$ in the TDGMT calculation ((a) TDGMT $-$ Born), $ph$-
contributions in the RPA approximation ((b) RPA $-$ Born) and mixed
diagrams in case of the full theory ((c) NQCD $-$ RPA $-$ TDGMT $+$ Born).
Obviously the $pp$-contributions are especially sensitive to the
short-range part of the potential while the $ph$-terms
are strongly influenced by the long-range part of the interaction.
Hence the $\delta$-force leads to an unrealistically high contribution
of the ladder diagrams to the nucleon-nucleon correlations while
an extremly long-range potential overestimates the loop diagrams.
These results - which confirm intuitive
expectations - quantitatively show the separation of short- and long-range
correlations, which are mainly generated by $pp$, $hh$ collisions
and $ph$ interactions, respectively.

\zeqn
\section{Summary}

In this study we have solved numerically the coupled equations of motion
for the one-body density matrix and the two-body correlation function
in a stationary and/or time dependent oscillator basis as well as within
a time dependent Hartree-Fock basis for the dynamical response of light
nuclei. In the stationary limit we recover the Brueckner G-matrix theory
when neglecting $ph$ interaction matrix elements, or the familiar integral
equation for the polarization matrix when discarding higher order $pp$
and $hh$ interactions. The full theory - denoted by NQCD - thus includes
an infinite resummation of loop and ladder diagrams as well as their mixed
diagrams nonperturbatively. In addition to our earlier
studies we have explicitely examined the influence of the trace conserving
terms proposed in \cite{10} for more realistic 3-dim. finite systems and
carried out calculations for the first time with finite range two-body
interactions which allowed to explore - at variance to current TDDM
calculations \cite{15} - the sensitivity of dynamical
quantities with respect to long- and short-range correlations.

The role of long- and short-range correlations is studied in context with
different physical quantities of interest in nuclear dynamics.
The dynamical problems addressed include the redistribution of occupation
numbers due to correlations for configurations close to the groundstate,
the relaxation of single-particle excitations, the damping of isoscalar
giant monopole resonances and the sensitivity of the
ISGMR width to the incompressibility $K$ in finite nuclei.
In order to follow the time evolution of different correlation
components connecting, e.g. two $pp$-states, two $ph$-states and so on,
we have discriminated the two-body correlations with respect to
the Hartree-Fock groundstate as defined in eqs. (6.37) to (6.42).
Due to the different problems addressed we enumerate the basic results:

\begin{itemize}

\item{The two-body trace relation (2.8, 2.9) is dynamically conserved in time
when including the extra trace-conserving terms proposed in \cite{10} for
spin-symmetric systems. Their dynamical influence on the investigated
physical quantities, however, is found to be negligible in all cases.}

\item{The problem of short- and long-range correlations is traced back
to the question how the formation of correlations in the different
channels is influenced by the range of the two-body interaction. It turns
out that a long-range force mainly produces correlations in the $ph$-channel
(RPA), whereas a short-range force favours the $pp$-channel (TDGMT). This
result is in agreement with earlier discussions (see e.g. \cite{7d,7e})
and confirms the intuitive expectation that in-medium two-body collisions are
controlled by short-range interactions, whereas the collective response
originates essentially from long-range parts of the interaction.}

\item{The redistribution of single-particle occupation numbers is sensitive
to the strength of the correlations, however, not explicitely to long- or
short-range correlations. With zero-range interactions the $pp$-channel
clearly controls the redistribution
while for finite-range interactions the $ph$-channel dominates.}

\item{The relaxation of single-particle excitations is governed both by the
$pp$ and $ph$ interaction matrix elements for zero-range forces, whereas
we find a sizeable destructive interference from the mixed diagrams in case
of long-range interactions. Thus neither a resummation of ladder diagrams
nor a resummation of loop diagrams alone will yield a proper answer for the
single-particle relaxation times.}

\item{The damping of the isoscalar giant monopole resonance (ISGMR)
is found to be dominated by  strong $2p-2h$ correlations.
Its width is approximately independent on temperature and on the nuclear
incompressibility $K$ in the physically relevant range. Only when adopting
a very low or high incompressibility,  a significant change with temperature
can be observed.}

\item{All the above phenomena can qualitatively be understood in the following
simple picture  that emerges from the detailed microscopic studies: the
$2p-2h$ correlations - irrespective of the range of the interaction -
initiate the smoothing of the Fermi surface and in this way open phase space
for the other correlations. While the $2p-2p$ correlations are mainly
generated by the ladder and mixed diagrams in case of zero-range
interactions, one also has a sizeable contribution from the loop diagrams
in case of finite range interactions. The $2h-2h$ correlations remain
small in all cases whereas the $ph-ph$ correlations are almost exclusively
produced by loop diagrams. The doorway correlations of
$2p-ph$ or $2h-hp$ nature show a similar behaviour; they are dominantly
generated by loop and mixed diagrams.}

\end{itemize}

Inspite of the quantitative guideline given in this work with respect to the
partial applicability of ladder or loop resummations for single-particle
or collective nuclear responses, we  note that the influence of the
mixed diagrams in most cases cannot be disregarded and that the full theory
NQCD - possibly including the trace conserving interaction terms of higher
order - should be applied.

\vspace{2cm}

The authors acknowledge valuable discussions with F. V. De
Blasio, P. F. Bortignon, R. A. Broglia and H. C. D\"onges
on various aspects of the present
work.
\vspace{2cm}
\section*{Appendix}
\zeqn
\begin{appendix}
\section{Dispersion of one-body operators}
The dispersion of a one-body operator $O$ is given by
\bea
\sigma^2_{O} &=& \langle O^2 \rangle - \langle O \rangle ^2 =
\sum_{\alpha\beta\alpha'\beta'}
\langle\alpha |O| \beta\rangle \langle \alpha'|O| \beta'\rangle
\langle a_\alpha^\dagger a_\beta a_{\alpha'}^\dagger a_{\beta'} \rangle
-  \langle O \rangle ^2
\nonumber\\
&=& \sum_{\alpha\beta\alpha'\beta'}
\langle\alpha |O| \beta\rangle \langle \alpha'|O| \beta'\rangle
\lbrace  \delta_{\alpha'\beta}\langle a_\alpha^\dagger a_{\beta'} \rangle
+ \langle a_\alpha^\dagger a_{\alpha'}^\dagger a_{\beta'} a_{\beta}
\rangle \rbrace
 - \langle O \rangle ^2
\nonumber\\
&=& \sum_{\alpha\beta\beta'}
\langle \alpha | O | \beta \rangle \langle \beta |O| \beta'\rangle
\rho_{\beta'\alpha} +
\sum_{\alpha\beta\alpha'\beta'}
\langle \alpha |O| \beta \rangle \langle \alpha'|O|\beta'\rangle
\lbrace C_{\beta\beta'\alpha\alpha'}
- \rho_{\beta\alpha'}\rho_{\beta'\alpha} \rbrace.
\nonumber\\
\label{a1}
\eea
For the quadrupole operator $Q_2({\bf r})$, i.e.
\be
O = Q_2({\bf r}) = \frac{1}{2} \sqrt{\frac{5}{4 \pi}} (2z^2 -x^2 - y^2),
\label{a2}
\ee
this gives for the corresponding velocity
\bea
\dot{Q}_2 = \frac{i}{\hbar} [{\cal H}, Q_2] =
\sqrt{\frac{5}{4 \pi}} \frac{1}{2m} (2zp_z+2p_z z -xp_x-p_x x -y p_y-p_y y)  .
\label{a3}
\eea
Together with eq. (\ref{a1}) one can now determine $\sigma_{\dot{Q}_2}^2$.
In the limit $(t \rightarrow \infty)$ $\sigma_{\dot{Q}_2}^2$ can be replaced
by $\langle \dot{Q}_2^2 \rangle$.
This allows to relate the dispersion of the collective velocity
by the equipartition theorem to a temperature $T$ via
\be
\frac{1}{2} M_{Q_2} \sigma_{\dot{Q}_2}^2 = \frac{1}{2} \; T \;,
\label{a4}
\ee
where the collective mass parameter is calculated according to \cite{37}
\be
M_Q^{-1} = \frac{i}{\hbar} \sum_{\alpha\alpha'} \rho_{\alpha\alpha'}
\langle \alpha'| [\dot{Q},Q] |\alpha \rangle .
\label{a5}
\ee

for $Q= Q_2$ or $Q_0$, respectively.

%
\section{Equation of motion for the oscillator \protect\\parameter}
In order to study the ISGMR it is convenient to choose
a time-dependent harmonic oscillator basis.
This choice makes it necessary to modify the l.h.s. of the
equations of motion (\ref{2.1}) and (\ref{2.2}) according to
\bea
i \frac{\partial}{\partial t}\: \rho_{\alpha\alpha'} (t)
\longrightarrow
 i \frac{\partial}{\partial t}\: \rho_{\alpha\alpha'}(t)
+i \sum_{\lambda} \lbrace
\langle\alpha| \dot{\lambda}\rangle \rho_{\lambda\alpha'}(t) +
\rho_{\alpha\lambda}(t) \langle\dot{\lambda}|\alpha'\rangle \rbrace
\label{B.1}
\eea
\bea
i \frac{\partial}{\partial t}\: C_{\alpha\beta\alpha'\beta'}(t)
\longrightarrow
i \frac{\partial}{\partial t}\: C_{\alpha\beta\alpha'\beta'}(t)
+ i \sum_\lambda \lbrace
\langle\alpha|\dot{\lambda}\rangle C_{\lambda\beta\alpha'\beta'}(t)+
\langle\beta|\dot{\lambda}\rangle C_{\alpha\lambda\alpha'\beta'}(t)
\nonumber\\
+C_{\alpha\beta\lambda\beta'}(t)\langle\dot{\lambda}|\alpha'\rangle +
C_{\alpha\beta\alpha'\lambda}(t)\langle\dot{\lambda}|\beta'\rangle\rbrace.
\qquad\quad
\label{B.2}
\eea
The matrixelements $\langle\alpha|\dot{\lambda}\rangle$ and
$\langle\dot{\lambda}|\alpha'\rangle$ within the oscillator basis
are given by
\be
\langle\alpha|\dot{\lambda} \rangle =
\sum_{i=1}^{3} \left\lbrace
\langle \alpha_i |\dot{\lambda}_i \rangle
\prod^3_{j \ne i}
\delta_{\alpha_j \lambda_j}
\right\rbrace
\label{B.3}
\ee
with
\be
\langle \alpha_i|\dot{\lambda_i} \rangle = \frac{\dot{b}}{2b}
\left[ \delta_{\lambda_i+2,\alpha_i}
\sqrt{(\lambda_i+2)(\lambda_i+1)} -
\delta_{\lambda_i-2,\alpha_i}
\sqrt{\lambda_i (\lambda_i -1)}
\right] \;
\label{B.4}
\ee
where the numbers $\lambda_i, \alpha_i$ (i= 1, 2, 3) correspond to the
quantum numbers $n_x, n_y, n_z$ of the harmonic oscillator basis in
cartesian representation.

The time dependence of the basis can be obtained
by requiring energy conservation after the excitation of the
collective mode. In addition to the kinetic, mean-field and
correlation energy one has to take into account a collective
energy given by
\be
E_{{\cal C}oll}(t)=\frac{1}{2}\: M_{Q_0}(t)\: \dot{Q_0}^2(t)
\label{B.5}
\ee
with a collective velocity
\bea
\dot{Q_0}(t) &=& \frac{\partial}{\partial t}\:  \langle r^2\rangle
\:\approx 2\: \frac{ \dot{b}(t)}{b(t)} \langle r^2 \rangle
\label{B.6}
\eea
and a mass parameter obtained according to (\ref{a5})
\bea
M_{Q_0} \: = \frac{m}{4 \langle r^2\rangle}.
\label{B.7}
\eea
Hence the collective energy can be rewritten in the form
\be
E_{{\cal C}oll} (t) = \frac{1}{2}\; m \langle r^2 \rangle
\left(  \frac{ \dot{b}(t)}{b(t)} \right)^2
\label{B.8}
\ee
Finally we obtain two coupled differential equations
for $\dot{b}(t)$ and $b(t)$
\bea
\frac{\partial}{\partial t}\: \dot{b}(t) &=&
\frac{1}{{\cal C} (b(t))}
\left \lbrace  {\cal H}(b(t)) -
\frac{ {\cal G} (b(t)) }{ \dot{b}(t) } \right \rbrace
\label{B.9}\\
\nonumber\\
\frac{\partial}{\partial t}\: b(t) &=& \dot{b} (t) \;\;,
\label{B.10}
\eea
which determine the time evolution of the harmonic oscillator basis.
For abbreviation we have introduced
\bea
{\cal C} (b(t)) &=& m \: \frac{\langle r^2\rangle}{ b^2(t)} \qquad\qquad\qquad
\label{B.11}\\
\nonumber\\
{\cal G} (b(t)) &=& 4 \: \sum_{\lambda\lambda'}
\langle\lambda|t|\lambda'\rangle
\dot{\rho}_{\lambda\lambda'}(t)
\nonumber\\
&+& 2 \sum_{\lambda\lambda'\alpha\alpha'}
\langle\alpha'\lambda'|v|\alpha\lambda\rangle_{\cal A}
\left[ \dot{\rho}_{\lambda\lambda'}(t) \rho_{\alpha\alpha'}(t)
+ \rho_{\lambda\lambda'}(t)\dot{\rho}_{\alpha\alpha'}(t) \right]
\nonumber\\
&+& 8 \sum_{\lambda\lambda'\alpha\alpha'}
\dot{C}_{\alpha\lambda\alpha'\lambda'}(t)
\langle\alpha'\lambda'|v|\alpha\lambda\rangle\qquad\qquad\qquad
\label{B.12}\\
\nonumber\\
{\cal H} (b(t)) &=& 8\: \sum_{\lambda\lambda'} \langle\lambda|t|\lambda'\rangle
\frac{\rho_{\lambda\lambda'}(t)}{b(t)}
\nonumber\\
&-&2 \sum_{\lambda\lambda'\alpha\alpha'}
\left( \frac{\partial}{\partial b}
\langle\alpha'\lambda'|v|\alpha\lambda\rangle_{\cal A} \right)\:
\rho_{\lambda\lambda'}(t)\rho_{\alpha\alpha'}(t)
\nonumber\\
&-& 8 \sum_{\lambda\lambda'\alpha\alpha'}
C_{\alpha\lambda\alpha'\lambda'}\:
\left( \frac{\partial}{\partial b}
\langle\alpha'\lambda'|v|\alpha\lambda\rangle \right)
\nonumber\\
&-& \frac{m}{2}\: \sum_{\alpha\alpha'}
\dot{\rho}_{\alpha\alpha'}(t)
\langle\alpha'||r^2||\alpha\rangle \dot{b}(t)
\label{B.13}
\eea
with the time independent matrix element
\be
\langle\alpha'||r^2||\alpha\rangle \:
\equiv \frac{1}{b^2(t)} \langle\alpha'|r^2|\alpha\rangle .
\label{B.14}
\ee
\section{Interaction matrix elements}
The interaction matrix elements are given by
\bea
\langle \alpha\beta |v|\alpha'\beta'\rangle =
\int \psi^*_\alpha ({\bf r_1}) \psi^*_\beta ({\bf r_2})
v(|{\bf r_1 -r_2}|)
\psi_{\alpha'} ({\bf r_1}) \psi_{\beta'} ({\bf r_2})
d^3 r_1 d^3 r_2.
\label{a20a}
\eea
In cartesian coordinates the $3$-dim. harmonic oscillator wavefunctions
$\psi_\alpha ({\bf r_1})$ separate into $x$-, $y$- and $z$-components, i.e.
$\psi_\alpha ({\bf r_1}) = \Phi_{n^1_x} (x_1) \Phi_{n^1_y} (y_1)
\Phi_{n^1_z} (z_1).$
In order to reduce the computational effort ($6$-dim. integration
in eq. (\ref{a20a})) we exploit that the motion of $2$ particles
in a harmonic oscillator potential can be seperated into a relative and
a center of mass part.
The transformation to relative and center of mass coordinates can be
determined for each component independently:
\be
\Phi_{n^1_x}\left(\frac{x_1}{b}\right) \Phi_{n^2_x}\left(\frac{x_2}{b}\right) =
\sum_{N_x n_x} {\cal T}^{n^1_x n^2_x}_{N_x n_x}\quad
\Phi_{N_x}\left(\frac{X}{B}\right) \Phi_{n_x}\left(\frac{x}{\tilde b}\right)\;,
\label{a20}
\ee
where $X=1/2 (x_1 + x_2)$ and $x = (x_1 - x_2)$ denote the relative and
center of mass coordinates, respectively, while $B = b/\sqrt{2}$
and $\tilde b = \sqrt{2}\;b$ are the corresponding oscillator parameters.
The explicit transformation matrix ${\cal T}$ is given by (ref. \cite{31}):
\be
{\cal T}^{n^1_x n^2_x}_{N_x n_x} = \sqrt{\frac{N_x! n_x!}{n^1_x! n^2_x!}}
\;\left(\frac{1}{2}\right)^{\frac{1}{2}(n^1_x + n^2_x - n_x)}
\sum_{i,k} (-1)^k {n^1_x \choose i} {n^2_x \choose k}
\left(\frac{1}{2}\right)^{i+k}.
\label{a21}
\ee
The sums over $i$ and $k$ are restricted by
\be
0 \leq i \leq n^1_x \qquad  0 \leq k \leq n^2_x \qquad i+k=n_x \;,
\label{a22}
\ee
while energy conservation is ensured via the requirement
\be
n^1_x + n^2_x = N_x + n_x \;.
\label{a23}
\ee
The matrix ${\cal T}^{3dim}$ for the transformation of $3$-dim.
wavefunctions is then simply given by
\be
{\cal T}^{3dim} = \left\lbrace {\cal T}^{n^1_x n^2_x}_{N_x n_x}
   \times  {\cal T}^{n^1_y n^2_y}_{N_y n_y}
   \times  {\cal T}^{n^1_z n^2_z}_{N_z n_z}\right\rbrace .
\label{a24}
\ee
After this transformation is performed one can carry out the
integration with respect to the center of mass coordinates
using the orthonormality of the oscillator wavefunctions and the
dependence of the interaction only on the relative coordinates.
In order to determine the remaining $3$-dim. integral over the
relative coordinates we go over to a spherical representation.
The formal structure of the transformation is given by
\be
\Psi_{n l m} \left(\frac{{\bf r}}{b}\right) = \sum_{n_x n_y n_z}
{\cal A}(n l m ; n_x n_y n_z) \;
\Psi_{n_x n_y n_z}\left(\frac{x}{b},\frac{y}{b},\frac{z}{b}\right)
\label{a25}
\ee
and
\be
\Psi_{n_x n_y n_z}\left(\frac{x}{b},\frac{y}{b},\frac{z}{b}\right)=
\sum_{n l m} {\cal A}^{*}(n l m ; n_x n_y n_z)\;
\Psi_{n l m}\left(\frac{{\bf r}}{b}\right)\;,
\label{a26}
\ee
with
\be
{\cal A}^{*}(n l m; n_x n_y n_z) = (-1)^m\; {\cal A}(n l \: - \! m; n_x n_y
n_z).
\label{a27}
\ee
Explicitly the transformation matrix ${\cal A}$ reads (ref.\cite{32}):
\bea
\lefteqn{{\cal A}(nlm,n_x n_y n_z) = \delta_{n_x+n_y+n_z,2n+l}\;
{\cal M}_{(k+m)k} (n_1) \;(-1)^{n-m}\;
2^{\frac{1}{2} (2k+l-m)}}&&\nonumber \\
\nonumber\\
& &\times
\left[\frac{(2n-1)!!(2l+1)!!(l+m)!(l-m)!(l)!(2n+l-m-2k)!k!}{(2n+2l+1)!!
(2l)!(2n)!(k+m)!}\right]^{\frac{1}{2}} \nonumber\\
\nonumber\\
& &\times \sum_{\sigma\beta}\left\lbrace 2^{-(\sigma+\beta)}\; (-1)^{\sigma}
{l+k-\sigma \choose l} {2n+2\sigma-2k \choose \sigma-\beta}\right. \nonumber\\
\nonumber\\
& &\times\left.
{n \choose k-\sigma} {l-m-\sigma \choose \beta}
\frac{(k-\sigma)!}{(l-m-\sigma)!}
\right\rbrace\;,
\label{a28}
\eea
with
\bea
{\cal M}_{N_+N_-}(\alpha) &=& \frac{1}{\sqrt{N_+!N_-!}}\; 2^{-\frac{1}{2}
(N_++N_-)}
\sqrt{\alpha ! \;(N_+ + N_- - \alpha)!} \nonumber\\
&\times&i^{(N_+ - N_- - \alpha)}
\sum_{\beta} (-1)^{\beta} {N_+ \choose \alpha-\beta} {N_- \choose \beta}\;.
\label{a29}
\eea
For abbreviation we have used $k \equiv (n_1 + n_2 - m)/2$ with the
restricting requirement that $2k$ has to be an integer.
In eq. (\ref{a28}), furthermore, one has to use $(2n - 1)!! =1 $
in case of $n=0$.
The energy conservation is guaranteed by
$\delta_{(n_x + n_y + n_z),2n+l}$.
After performing the integration over the angular part the problem
is traced back to an $1$-dim. integration:
\be
<nl|v(r)|n'l> = \int\limits_{0}^{\infty} R^l_n(r)\; v(r)\; R^l_{n'}(r) \;dr\;,
\label{a30}
\ee
where $R^l_n(r)$ is the radial part of a harmonic oscillator
wavefunction in spherical coordinates.
By performing an expansion in terms of Talmi integrals (ref. \cite{29})
\be
<nl|v(r)|n'l> = \sum_{p} {\cal B}(nl,n'l;p)\; I_p \;,
\label{a31}
\ee
where the Talmi integrals are given by
\be
I_p = \frac{2}{\Gamma (p+\frac{3}{2})}
\int\limits_{0}^{+\infty} r^{2(p+1)}\: e^{-(\frac{r}{b})^2}\; v(r)\; dr
\label{a32}
\ee
one can further simplify the problem.
The expansion coefficients ${\cal B}$ explicitly reads (ref. \cite{33}):
\bea
\lefteqn{{\cal B}(nl,n'l;p) = (-1)^{p-l}\; \frac{(2p+1)!}{2^{n+n'}\; p!}
\left[ \frac{n!\:n'!\:(2n+2l+1)!\:(2n'+2l+1)!}{(n+l)!\:(n'+l)!}
\right]^{\frac{1}{2}}} & & \nonumber \\
& &\!\!\!\!\!\!\!\!\!\!\!\times
\sum_{k=k_0}^{k_1} \left\lbrace
\frac{(l+k)!\;(p-k)!}{k!(2l+2k+1)!(n-k)!(2p-2k+1)!(n'-p+l+k)!(p-l-k)!}\right\rbr
ace.\nonumber\\
\label{a33}
\eea
The summation in eq. (\ref{a33}) is restricted to
\be
k_0 = \left\lbrace {0 \qquad\qquad\qquad\! p-l-n' \leq 0 \atop p-l-n' \qquad
p-l-n'>0} \right.
\label{a34}
\ee
and
\be
k_1 = \left\lbrace { p-l \qquad\qquad\, p-l-n\leq 0 \atop n
\qquad\qquad\qquad\!
p-l-n>0}\right.\;.
\label{a35}
\ee
In case of the Yukawa interaction
\be
v(r) = \left(\frac{V_0\:\hbar c }{r}\right) e^{-\frac{m}{\hbar c} r}
\label{a36}
\ee
the Talmi integrals (\ref{a32}) can be rewritten in terms of repeated
integrals over the error function $i^n \erfc(z)$ (ref. \cite{33})
\be
I_p = V_0 \; \frac{2^{p+1}}{\sqrt{\pi}} \; p! \;
\left(\frac{\hbar c}{\tilde b}\right) \;
e^{\mu^2} \; i^{2p+1}\erfc(y) \;,
\label{a37}
\ee
where we have used the abbreviations $\mu = \tilde b \: m /(2\: \hbar c)$
with the $\tilde b = \sqrt{2}\; b$.
The repeated integrals over the error function $i^n \erfc(z)$ in
eq. (\ref{a37}) are defined according to ref. \cite{38}.
The presented transformation scheme allows to expand the
interaction matrix elements (\ref{a20a}) in terms of
analytically well known functions.

In order to get the derivatives of the matrix elements with respect
to $b$ - which are necessary for the coupled equations of motion for
$b(t)$ and $\dot{b}(t)$ - one has to determine the derivative of
(\ref{a37}) with respect to $b$.
With
\be
\frac{d}{d\mu} \; i^n \erfc(\mu) = -i^{n-1} \erfc(\mu) \qquad (n=0,1,2,..)
\label{a38}
\ee
one obtains
\bea
\frac{d}{db}\;I_p = 2^{2p}\; p!\; \left( \frac{\hbar}{c}\right)\; e^{\mu^2}\,
\left\lbrace -\frac{\sqrt{2} \; i^{2p+1} \erfc(\mu)}{b^2}
+ \right.\qquad\qquad\nonumber\\ \left.
\frac{m(b_0)}{b\, \hbar c} \left[
1+\alpha\left(1+2\;\frac{b_0^3}{b^3}\right)\right]
\left[ 2\mu \; i^{2p+1} \erfc(\mu) - i^{2p} \erfc(\mu) \right] \right\rbrace
\;\; ,
\label{a39}
\eea
where a linear density dependence of the meson masses according
to eq. (\ref{5.4}) was assumed.
\end{appendix}
%
%
%
%

%
%
\section*{Figure captions}

Fig. 1: Asymptotic fluctuations in the quadrupole velocity
$\sigma^2_{\dot{Q_2}} (\infty)$ multiplied by the collective
mass parameter $M_{Q_2}$ as a function of the initialization temperature
$T$. The straight line represents the values expected according
to eq. (\ref{a4}) in line with the equipartition theorem.
\vspace{0.9cm}

Fig. 2: The quantity $H_{diff}(t)$ for the limites BORN,
TDGMT, NQCD ($\Delta t = 0.5 \times 10^{-23}$s) and
SCD ($\Delta t = 0.5$, $0.4\times 10^{-23}$s).
In the insertion the time evolution of $H_{diff}(t)$ in the SCD limit
is displayed on a longer time scale.
\vspace{0.9cm}

Fig. 3: Distribution of occupation numbers close to the
groundstate for the TDDM approach in case of $^{16}O$ (a) and
$^{40}Ca$ (b) in comparison with our model taking into account
the $\delta$-force and the Yukawa potential. Experimental values
from ref. \cite{18} are indicated for comparison.
In addition the hole strength as arising from our model approach
is displayed within different approximations for the $\delta$-force
(c) and the Yukawa interaction (d).
\vspace{0.9cm}

Fig. 4: Time evolution of the occupation probability $n_{\lambda}(t)$
within different approximations for the $\delta$-force.
The dotted lines indicate the average occupation after equilibration
and the least-square fit with the trial function (\ref{4.1}).
\vspace{0.9cm}

Fig. 5: Damping width of the single-particle excitation
within different approximations. For comparison we display
the values obtained with the $\delta$-force and the Yukawa
potential, respectively.
\vspace{0.9cm}

Fig. 6: Equation of state (EOS) for our model $^{16}O$-nucleus for
different density dependences $\alpha_{DD}$ of the exchange mesons. Increasing
values of the parameter $\alpha_{DD}$ are related to an increasing
value of the incompressibility $K$ (solid lines).
The relevant parts of the EOS are indicated in case of 15 and
30 MeV excitation energy (dotted lines).
The small figure displays the incompressibility $K$ (solid line) and
$K_\infty = 3/2 K$ (dashed line) as a function of $\alpha_{DD}$.
Commonly accepted values of $K_\infty$ and the corresponding parameters
$\alpha_{DD}$ are indicated by dotted lines.
\vspace{0.9cm}

Fig. 7: The model potential consisting of an attractive
$\pi$-exchange and a repulsive $\omega$-exchange term for
$\alpha_{DD}=0.30$ and $\rho = 0.5\rho_0 $ (solid line),
$\rho=\rho_0 $ (dashed line) and $\rho =2 \rho_0$ (dotted line).
The small figure shows the lowest oscillator
wavefunction $\Phi_0$ for the same three values of $\rho$.
\vspace{0.9cm}

Fig. 8: Density oscillation ($\rho/\rho_0 = b_0^3/b^3$) of the
model $^{16}O$-nucleus in TDHF (solid lines) and NQCD (dotted lines)
for 15 and 30 MeV excitation energy, $T=0$ MeV temperature
and $\alpha_{DD}$ = 0.0 and 0.3.
\vspace{0.9cm}

Fig. 9: Time evolution of the monopole moment $Q_0(t)$ for the model
$^{16}O$-nucleus ($T=1$ MeV, $\alpha_{DD}=0.1$)
which was excited at $t_0 = 18, 23, 28$ and $33\times 10^{-23}$s
with 15 MeV and 30 MeV collective energy, respectively. The time scales
are shifted to $t_0 = 0$.
\vspace{0.9cm}

Fig. 10: Monopole moment $Q_0 (t)$ for the model $^{16}O$-nucleus
($T=1$ MeV, $\alpha_{DD}=0.2$) within different
possible approximations on the two-body level.
\vspace{0.9cm}

Fig. 11: Damping width of the ISGMR as a function of the temperature $T$
in NQCD approximation for different values of $\alpha_{DD}$ and
$15$ MeV excitation energy.
\vspace{0.9cm}

Fig. 12: $Q_0 (t)$ and $S_0 (E)$ for the model $^{16}O$-nucleus
($T=0$ MeV) for different values of the parameter
$\alpha_{DD}$ within the NQCD approach. The excitation energy is $15$ MeV.
\vspace{0.9cm}

Fig. 13: Damping width of the ISGMR as a function of the parameter
$\alpha_{DD}$ within the NQCD approach for $15$ and $30$ MeV
excitation energy. Physical values of the parameter $\alpha_{DD}$
are indicated by the dotted lines.
\vspace{0.9cm}

Fig. 14: Time evolution of the different types of correlations
for the model $^{16}O$ nucleus initialized at $T=0$ MeV
for different approximations in case of the $\delta$-interaction.
\vspace{0.9cm}

Fig. 15: Time evolution of the different types of correlations
for the model $^{16}O$ nucleus initialized at $T=0$ MeV
for different approximations in case of the Yukawa-interaction.
\vspace{0.9cm}

Fig. 16: Time-averaged distribution for the different types of correlations
within the NQCD approach for the $\delta$- and Yukawa interaction.
The normalization is carried out with respect to the sum over all
matrix elements $C_{\alpha\beta\alpha'\beta'}$.
\vspace{0.9cm}

Fig. 17: Time-averaged distribution of the different types of
correlations for T = 0, 2 and 4 MeV within the NQCD approach in
case of the Yukawa potential. The normalization is carried out with
respect to the sum over all matrix elements $C_{\alpha\beta\alpha'\beta'}$.
\vspace{0.9cm}

Fig. 18: Time-averaged sum over the matrix elements
$C_{\alpha\beta\alpha'\beta'}$ for different possible approximations.
The normalization is carried out with respect to the full theory (NQCD).
Ladder diagram contributions in the TDGMT calculation (a), loop-diagram
contributions in the RPA approach (b) and mixed contributions
in the NQCD calculation (c) are indicated as well.
\vspace{0.9cm}
\end{document}